\documentstyle[epsf]{mn}

\def\ltsim{\raisebox{-.5ex}{$\;\stackrel{<}{\sim}\;$}}

\def\vFWHM{\ifmmode V_{\mbox{\tiny FWHM}} \else
            $V_{\mbox{\tiny FWHM}}$\fi}
\def\kms{\ifmmode {\rm km\ s}^{-1} \else km s$^{-1}$\fi}
\newcommand{\et}{et al.\ }

\newcommand{\xte}{{\it RXTE}}

\title{Complex Optical--X-ray Correlations in the Narrow-Line Seyfert 1 Galaxy NGC~4051}

\author[O. Shemmer \et]
       {O.\ Shemmer,$^{1}$\thanks{ohad@wise.tau.ac.il} P.\ Uttley,$^{2,3}$ H.\ Netzer$^{1}$ and I.~M.\ M$\rm ^{c}$Hardy$^{2}$ \\
       $^1$School of Physics and Astronomy and the Wise Observatory, The
Raymond and Beverly Sackler Faculty of Exact Sciences, \\ Tel-Aviv
University, Tel-Aviv 69978, Israel \\
       $^2$Department of Physics and Astronomy, University of
		Southampton, Southampton SO17 1BJ, United Kingdom \\
       $^3$Visiting astronomer, Tel-Aviv University, Wise
		Observatory \\
}
\date{Accepted 2003 May 7.
      Received 2002 December 13;
      in original form 2002 December 13}
\pagerange{\pageref{firstpage}--\pageref{lastpage}}
\pubyear{2003}

\begin{document}

\maketitle

\label{firstpage}

\begin{abstract}

This paper presents the results of a dense and intensive X-ray and optical
monitoring of the narrow-line Seyfert 1 galaxy NGC~4051 carried out in 2000. Results
of the optical analysis are consistent with previous measurements.
The amplitude of optical emission line variability is a factor of two larger than
that of the underlying optical continuum, but part or all of the difference
can be due to host-galaxy starlight contamination or due to the lines being driven
by the unseen UV continuum, which is more variable than the optical continuum. We
measured the lag between optical lines and continuum and found a lower, more accurate
broad line region size of $3.0\pm1.5$ light days in this object. The implied black hole
mass is M$_{BH}=5^{+6}_{-3}\times10^5 M_{\odot}$; this is the lowest mass found, so far, for
an active nucleus. We find significant evidence for an X-ray--optical (XO) correlation
with a peak lag $\ltsim1$ day, although the centroid of the asymmetric correlation
function reveals that part of the optical flux varies in advance of the X-ray flux by
$2.4\pm1.0$ days. This complex XO correlation is explained as a possible combination of
X-ray reprocessing and perturbations propagating from the outer (optically emitting)
parts of the accretion disc into its inner (X-ray emitting) region.

\end{abstract}

\begin{keywords}

galaxies: active -- galaxies: individual: NGC~4051 -- X-rays: galaxies

\end{keywords}

\section{Introduction}
\label{intro}

Correlations between different parts of the spectral energy distribution
in active galactic nuclei (AGN), have been utilized in the past decade as
an important tool to probe and map the deepest components of the central engine's
energy source. Several attempts have aimed at finding a connection between X-ray
and optical light curves, in order to follow the source and location of the X-ray
emission. Such attempts have been carried out in the past decade as part of AGN
multiwavelength monitoring campaigns (e.g. Done \et 1990, Nandra \et 1998, 2000,
Edelson \et 2000, Maoz, Edelson \& Nandra 2000, Peterson \et 2000;P00, Shemmer \et
2001; see also Maoz \et 2002 for a brief summary of previous campaigns). So far,
reliable determinations of X-ray--optical (XO) correlations are few and far between.
In most cases where a strong correlation was found, the X-ray and optical light curves
appeared to vary simultaneously, i.e. practically with zero lag. For example, on long
timescales (days--months) all attempts to find XO lags have failed (e.g. Clavel \et 1992
in NGC~5548, Done \et 1990 and P00 in NGC 4051, Shemmer \et 2001 in Ark 564, and Maoz
\et 2002 in NGC 3516). Even on shorter timescales (hours) XO correlations and lags are
rare (e.g. Edelson \et 1996 in NGC~4151, but see also Edelson \et 2000 for
no XO correlation in NGC~3516).

One exception is NGC~7469 (Nandra \et 1998, 2000), in which a significant
correlation was found between the optical/UV continuum and the X-ray flux that followed
it with a $\sim4$ days lag, including periods when increasing X-ray flux led decreasing
UV flux by a similar lag. This complex behaviour ruled out two possible scenarios:
UV seed photons that are Compton up-scattered to produce X-rays in a putative
corona (UV leading X-ray; e.g. Haardt \& Maraschi 1991, 1993) or UV radiation that is
produced by reprocessed X-ray photons (X-ray leading UV; e.g. Stern
\et 1995).  Uttley \et (2003) find a very strong correlation between long
time-scale (months) X-ray and optical variations in NGC~5548, but constrain any
lag to be less than 15~days. Another success in finding an XO correlation was during the
Ark 564 campaign, when an X-ray flare was followed $\sim2$ days later by an optical flare
(Shemmer \et 2001) and was interpreted in terms of reprocessing models.

The successful detection of an optical response to an X-ray flare in Ark 564,
a narrow-line Seyfert 1 (NLS1) galaxy, has motivated us to search for
similar behaviour in other NLS1s that we have been monitoring as part of a
larger project. Since one of the more pronounced characteristics of NLS1s
is the intense X-ray variation (e.g. Boller, Brandt, \& Fink 1996;
Leighly 1999a,1999b), which is at least one order of magnitude
larger in amplitude than in `normal' Seyfert 1 galaxies, we assumed
that detection of XO connections will be more frequent and more pronounced
in this sub-class of AGN. One difficulty though, appears to be the fact
that the persistent large and rapid X-ray variability in NLS1s (flux variations
of a factor of two or more on timescales of minutes/hours) is contrasted by the
very low variability exhibited by the optical band. NLS1s differ markedly
from `normal' broad-line Seyfert 1 galaxies (e.g. NGC~7469; Nandra \et 1998, 2000)
in this respect by varying strongly in the X-ray while showing little or no variability
in the optical/UV band (e.g. Ark~564; Shemmer \et 2001).

NGC~4051 is a nearby ($z=0.0023$), low luminosity ($\sim10^{42} \ {\rm erg} \ {\rm s}^{-1}$), NLS1
(FWHM(H$\beta$)=1110 \kms) that had been studied extensively across the spectrum
(e.g. Uttley \et 1999, Lamer \et 2002, Collinge \et 2001 and references therein) and has shown
optical variability amplitudes of up to $\sim10\%$ in flux (Done
\et 1990, P00). In 2000 we carried out a dense and continuous X-ray and optical
monitoring campaign on NGC~4051. Our major goal aimed at finding a temporal
relationship between the variations observed in the two bands. In this paper we
present the results of this campaign. Section~\ref{observations}
presents the observational data and their reduction. In \S~\ref{results}
we present the results of the time series analysis and in \S~\ref{discussion}
discuss its implications. Section~\ref{conclusions} summarizes our main conclusions.

\section{Observations and Data Reduction}
\label{observations}
\subsection{The Optical Band}
\label{observations.1}

\begin{figure}
\centerline{\epsfxsize=3in\epsfbox{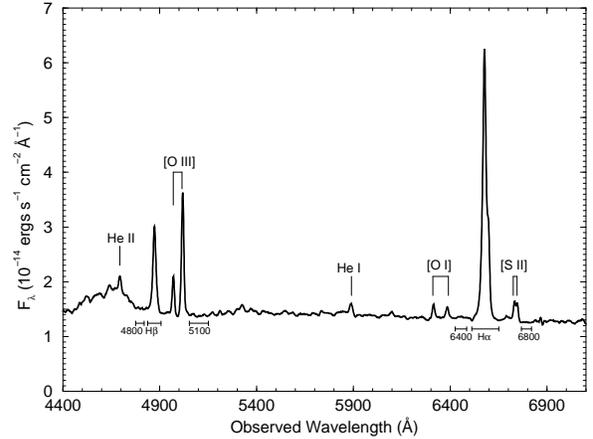}}
\caption{Mean spectrum of NGC 4051 observed at WO. The continuum and line
measurement bins are marked.}
\label{spectrum}
\end{figure}

\begin{figure}
\centerline{\epsfxsize=3in\epsfbox{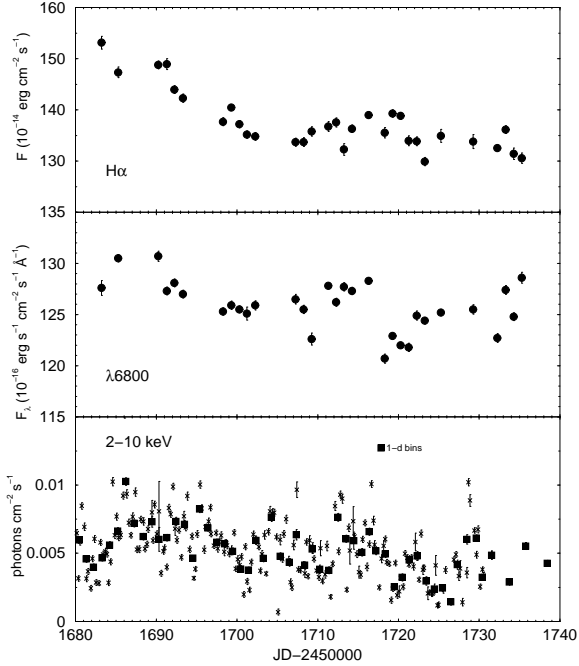}}
\caption{NGC 4051 light curves. From top to bottom: H$\alpha$ flux,
continuum flux density at the narrow 6800\AA \ band,
and \xte~2--10 keV flux (small 'x's with error bars). The X-ray data were also binned
in one-day intervals (filled squares).}
\label{lc}
\end{figure}

NGC~4051 was monitored spectrophotometrically during May--July 2000 at
the Tel-Aviv University Wise Observatory (WO). The observations were
carried out with the Faint Object Spectrograph \& Camera on top of the
WO 1m telescope. We used a 10''-wide long-slit and a 600 lines mm$^{-1}$
grism. A Tektronix 1024$\times$1024 pixel back-illuminated CCD was used
as the detector. Reduction of the data was carried out in the usual manner
using IRAF\footnote{{IRAF (Image Reduction and Analysis Facility) is distributed
by the National Optical Astronomy Observatories, which are operated by AURA, Inc.,
under cooperative agreement with the National Science Foundation.}} with its
{\sc{specred}}, {\sc{onedspec}} and {\sc{twodspec}} packages.
In order to reduce light contamination from the host galaxy while not lowering
the S/N ratio, we extracted the spectrum using an 8'' extraction window.
Spectrophotometric calibration of the nucleus of NGC 4051 was carried
out using the technique in which a nearby comparison star is observed
simultaneously with the object of interest inside a wide slit.
This technique of using a local comparison star is described in detail
by Maoz \et (1990, 1994) and produces high relative spectrophotometric accuracy.
Each spectroscopic observation consisted of two 15
minute exposures of NGC 4051 and its comparison star. The consecutive
galaxy/star flux ratios were compared to test for systematic errors in
the observations and to clean cosmic rays. We discarded pairs of data
points with ratios larger than $\sim$5\% and verified that the
comparison star is non-variable to within $\sim$2\% by means of
differential photometry of other stars in the field, carried out before
this campaign began. As a result, 31 good-quality spectra remained.
The spectra were calibrated to an absolute flux scale by multiplying
each galaxy/star ratio by a spectrum of the comparison star that was flux
calibrated by applying a characteristic WO extinction curve and CCD sensitivity
function, that do not change considerably from night to night. The absolute flux
calibration has an uncertainty of $\sim$10\%, which is not shown in the
error bars of our light curves. The error bars reflect only the
differential uncertainties, which are of order 2\%-3\%. By measuring
the [O {\sc iii}]$\lambda$5007 fluxes in our spectra, we verified
that the differential uncertainty level is consistent with the night-to-night scatter
in this narrow emission line light curve (which is expected
to maintain a constant flux level).
We measured the mean flux in narrow line-free continuum bands close to
H$\alpha$ and H$\beta$ (see Figure~\ref{spectrum}) and the integrated
flux of both emission lines in each spectrum. Two of the resulting light
curves (together with the X-ray light curve, see
\S\S~\ref{observations.2}) are plotted in Figure~\ref{lc}.

\subsection{X-ray Observations}
\label{observations.2}

NGC~4051 was intensively monitored by the {\it Rossi X-ray Timing Explorer} ({\it
RXTE}) in May--July 2000, as part of an ongoing campaign to measure its broad band
X-ray variability power spectrum (M$^{\rm c}$Hardy et al., in preparation). The
intensive monitoring program consisted of 251 observations, each of exposure
$\sim1$~ks, obtained at roughly 6-hourly intervals from 2000 May 1 to July 5.
We used data from the {\it RXTE} Proportional Counter Array, applying standard
good time interval selection criteria and using all available Proportional Counter
Units (PCUs; top layer only) to extract a spectrum for each observation.
Using {\sc xspec}, we fitted each spectrum with a simple power law plus Galactic
absorption model, in order to obtain an estimate of the 2--10~keV photon flux which
is robust to changes in instrument gain and number of PCUs used (see Lamer \et 2002
for further details of data reduction and spectral fitting).

\section{Time Series Analysis}
\label{results}

\subsection{Variability}
\label{results.1}

The fractional variability, $F_{var}$, (Rodriguez-Pascual \et 1997)
of a light curve is defined as
\begin{equation}
\label{eq:defF_var}
F_{var} = \sqrt{S^2 - \langle \sigma_{err}^2 \rangle \over \langle
X \rangle^2},
\end{equation}
where $S^2$ is the total variance of the light curve, $\sigma_{err}^2$
is the mean error squared, and $\langle X \rangle^2$ is the mean flux
squared. The uncertainty on $F_{var}$ is (Edelson \et 2002):

\begin{equation} \label{eq:defsigmaF_var}
\sigma_{F_{var}} = \frac{S^2}{\sqrt{2N}F_{var}{\langle{X}\rangle}^2}.
\end{equation}

\begin{table}
%\centering
%\begin{minipage}{140mm}
\caption{Fractional Variability \label{F_var}}
\begin{tabular}{@{}lr@{}}
Band & $F_{var} [\%]$ \\
2--10 keV & $44.9\pm2.0$ \\
$\lambda$4800 & $3.4\pm0.5$ \\
H$\beta$ & $7.8\pm1.2$ \\
%\protect{[O {\sc iii}]$\lambda$5007} & $2.5\pm0.6$ \\
$\lambda$5100 & $3.4\pm0.5$ \\
$\lambda$6400 & $1.8\pm0.3$ \\
H$\alpha$ & $4.0\pm0.5$ \\
$\lambda$6800 & $1.9\pm0.3$ \\
\end{tabular}
%\end{minipage}
\end{table}

A list of $F_{var}$ values calculated for the optical and X-ray light curves
appears in Table~\ref{F_var}, where it is apparent that the X-ray variability
is an order of magnitude larger than the optical variability over the same time
interval\footnote{Even though the sampling patterns of both X-ray and optical light curves
are different, both have a similar length and
a similar exposure time ($\sim10$ min.) for each data point. This allows us to directly
compare the $F_{var}$ values of the unbinned X-ray light curve to that of the optical.
The $F_{var}$ of the 1-d bin X-ray light curve is $35.3\pm3.1$\% and even though it might
seem more intuitive to compare this to the optical value, it is technically
incorrect since binning the X-rays smooths out the $\ltsim1$d variations that
contribute to the optical light curve.} It is also apparent that within the optical band
itself, the emission line variations are about twice as large as the underlying optical
continuum. This discrepancy may be attributable to contamination of the spectrum by
host-galaxy starlight, that tends to reduce the apparent optical continuum fluctuations.
For example, Done \et (1990) estimated respective $B$ and $I$ band contributions due to
NGC~4051 host-galaxy starlight of 33\% and 37\% of the observed continuum, for a 6\arcsec
aperture. Since we use a larger aperture (10\arcsec slit width and 8\arcsec extraction
region perpendicular to that), the contribution due to host galaxy starlight
may approach the 50\% level required to explain the factor of $\sim2$ difference in
line and continuum variability amplitudes. We tested this by analyzing images of the
galaxy taken in the $B$ band at WO. By scaling a PSF of other stars in
the galaxy's field to the galaxy's nucleus, and then subtracting it from the nucleus, we
found that the difference in flux is 15\% in a 10\arcsec \ by 8\arcsec aperture.
In other words, the starlight flux contribution to our spectroscopic aperture is at least 15\%,
which is not a meaningful lower limit due to the limited seeing conditions.
To ultimately test the `host-galaxy starlight contamination' scenario it is better
to examine UV continuum variations, since those are free from such contamination.
Examination of archival {\sl IUE} spectra taken between 1978 and 1994 shows periods of
large and long-term (years) UV variability, with amplitudes that are much larger than
those of the optical emission lines. However, the very large errors in the data prevent
us from obtaining $F_{var}$ for the UV continuum. We conclude that in spite of our
inability to quantitatively point at the source for the larger optical emission line
variability, fluctuations in the UV continuum, which are the most likely drivers of those
variations, remain a probable explanation that should be further checked with better
data than we currently have.

\subsection{Cross-Correlations}
\label{results.2}
\begin{figure*}
\centerline{\epsfxsize=4.5in\epsfbox{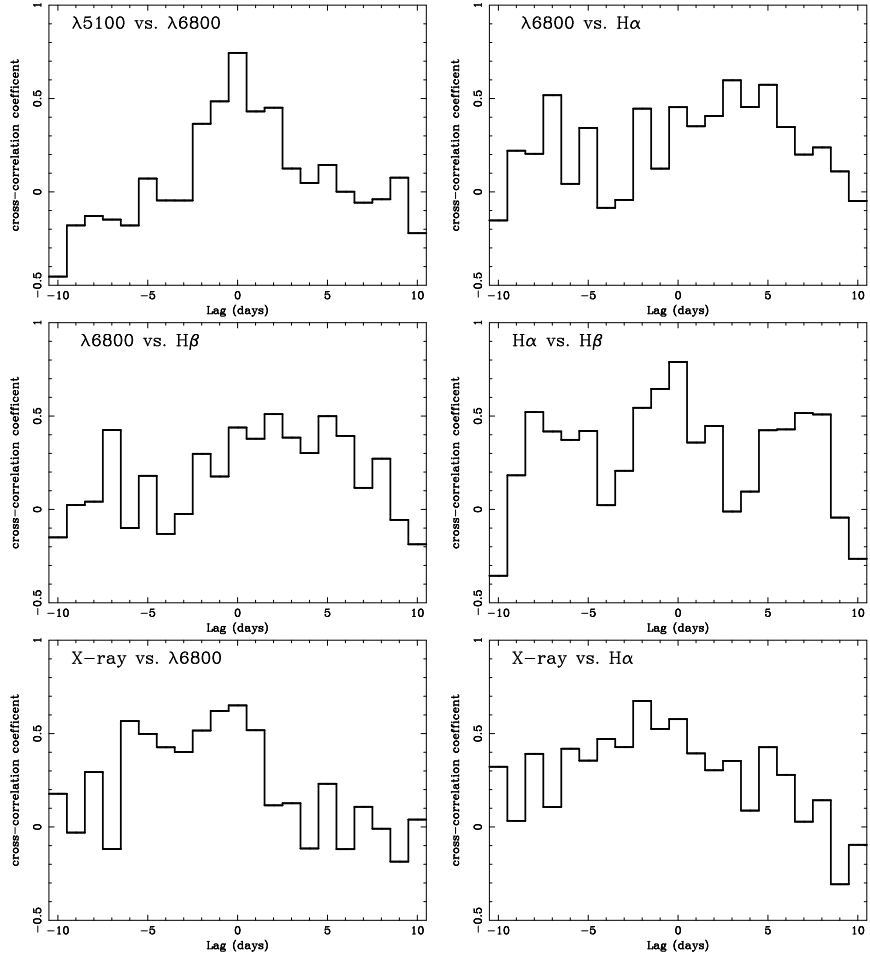}}
\caption{X-ray and optical correlations (see Table~\ref{ccf_results}
for parameters of the correlations). Positive lags imply the second light
curve lags the first. Note that although error bars are commonly plotted
on DCFs, we do not plot them here as they are meaningless for red-noise data,
since errors in the DCF are not independent.}
\label{ccffig}
\end{figure*}

%\subsection{X-ray and Optical Correlations \label{results.2}}
The X-ray light curve shows strong short time-scale variations which are not
reflected in the optical light curves (as previously noted by Done \et 1990 and P00),
so to obtain a better comparison with the optical variations we smoothed out the rapid
X-ray variability by binning the X-ray light curve into 1 day bins (see
Figure 2) before cross-correlating with the unbinned optical data (which is sampled at
$\sim$daily intervals). To derive the cross-correlation function (CCF) between
two light curves (the first assumed to be the driving light curve and the second
assumed to be the responding light curve) we utilized the Discrete Correlation
Function (DCF) method (Edelson \& Krolik 1988). For each pair of light curves,
we measured the DCF in the lag range $0\pm10$~d, binning in 1 day lag bins
(at larger lags there are fewer pairs of light curve points per lag bin so that
spurious peaks in the CCF are much more common). Peak values ($r_{\rm
max}$) and corresponding peak lags ($\tau_{\rm peak}$) were determined, together
with the lag centroid ($\tau_{\rm cent}$), which is a measure of the `centre of mass'
of the lag peak, and thus takes account of asymmetries in the correlation.
The centroid is determined by summing the CCF values in the range either side of
$\tau_{\rm peak}$ where the CCF value $r>0.8r_{\rm max}$. The uncertainties on
the lags (peak and centroid) were estimated using the Flux Randomization/Random
Subset Selection (FR/RSS) method (Peterson \et 1998). The CCFs for the most important
correlations are plotted in
Fig.~\ref{ccffig} and parameters of each correlation are shown in
Table~\ref{ccf_results}.

At face-value, the correlations shown in Table~\ref{ccf_results}
appear to be significant, with correlation coefficients larger than
one would expect if the data were randomly distributed and
uncorrelated.  However, the light curves presented here are not random
(white-noise) data sets: adjacent data points are correlated with one
another to produce variations on a range of time-scales and are
consistent with red-noise processes.  As such, the correlations
between two light curves are driven by only a few events (flares or
dips) in each light curve, and it is possible that apparent
correlations could be seen even where none exist, simply because the
events in two uncorrelated light curves happen to match up by chance.
To assign a reliable significance to the correlations, we must
simulate uncorrelated red-noise light curves with similar variability
properties to the observed light curves, and determine the frequency of
spurious correlations. A similar Monte Carlo method to assess the significance of
the XO correlation in NGC~5548 has been applied by Uttley \et (2003).
We outline the method here:
\begin{itemize}
\item[1.] Simulate two continuous red-noise light curves, of time resolution 0.01
days and length 16384 bins (i.e. 163~d, much larger than the 60~d observed
duration) using the method of Timmer \& Koenig (1995), with different random
number sequences to generate each light curve so they are uncorrelated. We
assume broken power-law shapes for both optical and X-ray power
spectra, with break frequencies at 1~d$^{-1}$ and power-law slopes
above the break of -1.5 and -2 for X-ray and optical power spectra
respectively and identical slopes of -1 below the break.  The X-ray
power-spectral shape is chosen to approximate that measured by much
more extensive {\it RXTE} and {\it XMM} data sets (M$^{\rm c}$Hardy et al.,
in preparation), while the optical power-spectral shape (which is
assumed to be the same for continuum and lines) is chosen to reproduce
the relatively low variability on short time-scales and mimic the
finding in NGC~5548 that the optical and X-ray power spectral shapes
differ only at high frequencies (Uttley et al. 2003)\footnote{To
ensure that our significance estimates are not strongly dependent on
optical power-spectral shape, which is ill-defined, we also tested
light curves with break frequency as low as 0.01~d$^{-1}$ and slope
above the break as steep as -2.5, and find no significant deviation
from the estimates we present here.}. Power-spectral amplitudes are
chosen so that the integrated power of the underlying power spectrum
gives the observed light curve variance, after observational noise is
subtracted (see Uttley, M$^{\rm c}$Hardy \& Papadakis 2002 for
further discussion of light curve simulation, as applied to the
measurement of power spectra).
\item[2.] Apply observational noise to the simulated light curves, by adding
to each simulated data point a random deviate of mean zero and variance equal
to the average squared error of the corresponding light curve.
\item[3.] Resample the simulated light curves to the observed sampling
patterns and rebin the simulated, resampled X-ray light curve to 1~d bins.
\item[4.] Measure the DCF of the pair of simulated, uncorrelated light
curves, and search for a peak value, $r_{\rm max}$, as outlined above
for the observed light curves. Search within lags of $\pm10$ days, allowing
adequate overlap between the two light curves.
\item[5.] Repeat steps 1-4 1000 times, and count the number of times that the
simulated $r_{\rm max}$ exceeds the observed $r_{\rm max}$, to yield the
significance of the observed correlation.
\end{itemize}
For example, for the X-ray--$\lambda$6800 correlation, we observed a maximum
correlation coefficient of $r_{\rm max}=0.65$.  We counted 40 out of 1000
simulated, uncorrelated light curves with $r_{\rm max}>0.65$, implying that
the observed correlation is significant at the 96\% confidence level
(i.e. just over 2$\sigma$). The estimated significance of each correlation
shown in Table~\ref{ccf_results}, is based on such simulations.

We note that our Monte Carlo approach shows that the actual significance
of the correlations is considerably less than would be expected from
white-noise data given the same values of $r_{\it max}$.  However, all the
optical continuum--continuum and line--line correlations are significant at
better than 95\% confidence, as are the XO correlations.  The
optical continuum--line correlations are not significant, but this
represents a limitation of the existing data set which contains few
events in each light curve (and some additional scatter which weakens the
correlation).  Longer data sets confirm that the optical continuum-line
correlation is real (Peterson \et 2000).  We stress however that the
questions of the significance of a correlation, and the significance of
lags measured between two light curves are not the same.  The
significance of a correlation can be determined by testing the
null-hypothesis that the light curves are uncorrelated.  However, in
order to determine the significance of any lag, one must assume that the
light curves are indeed correlated (as is implicit in the FR/RSS method of lag
error estimation); in that case the quality of sampling is important to
constrain the lag, rather than the number of `events' in the light curve.
Therefore we can still measure lags which are well constrained, even
though the correlation itself is not formally significant.

\begin{table*}
\centering
\begin{minipage}{140 mm}
\caption{NGC 4051 Cross-Correlation Results. \label{ccf_results}}
\begin{tabular}{@{}llrrr@{}}
Bands                          & $r_{max}$ &  Significance of $r_{max}$ & $\tau_{peak}$ (days)  &  $\tau_{cent}$ (days) \\
5100\AA--6800\AA               & 0.74      &          0.96              & $0\pm1$        & $0.4\pm0.9$    \\
6800\AA--H$\alpha$             & 0.60      &          0.85              & $3.0\pm1.5$    & $3.1\pm1.6$    \\
X--6800\AA                     & 0.65      &          0.96              & $0\pm1$        &  $-2.4\pm1.0$  \\
X--H$\alpha$                   & 0.67      &          0.97              & $-2\pm3$       & $-0.9\pm1.7$   \\
6800\AA--H$\beta$              & 0.51      &          0.69              & $2.0\pm2.6$        & $2.0\pm2.3$        \\
H$\alpha$--H$\beta$            & 0.79      &          0.98              & $0\pm2$        & $-0.9\pm1.7$        \\
\end{tabular}
\end{minipage}
\end{table*}

\section{Discussion}
\label{discussion}

We have monitored NGC 4051 in X-ray and in the optical band on a daily
basis for about 60 days in order to find a possible relation between
the two bands. Our main observational results
are discussed below.

\subsection{Optical Line--Continuum Lag}
\label{discussion.2}

Cross-correlations between the two major Balmer emission-lines and the
optical continuum confirm the previously detected lag in this object
(P00). We find that H$\alpha$ responds to the continuum variations
after $3.0\pm1.5$ days, which is consistent, within the errors, with the
$5.92^{+3.13}_{-1.96}$ days reported in P00 for H$\beta$. Since we do not
detect any lag between H$\alpha$ and H$\beta$, our new line--continuum lag
has a lower error, perhaps due to the denser sampling frequency (about once a day)
compared with the previous campaign (about once every four days; P00). Moreover,
as the observed average flux of NGC~4051 in this study is similar
(to within $\sim10\%$) to that observed during all three phases
of the P00 campaign, we suggest that the lower lag we find is
not a luminosity effect, but the combined effect of observations
and the CCF. By incorporating our lowest error value for the lag
($3.0\pm1.5$ days), and FWHM(H$\beta$)=1110$\pm$190 \kms \ from P00
into Eq. 5 of Kaspi \et (2000) for the virial black hole (BH) mass estimate,
we obtain M$_{BH}=5^{+6}_{-3}\times10^5 M_{\odot}$. Our result is thus
consistent, within the errors, with the P00 estimate. The new and lower broad line region
(BLR) size we obtained, $R_{BLR}=3.0\pm1.5$ light days, places NGC~4051
much closer to the best-fit $R_{BLR}$--$L$ slope produced from reverberation
measurements of 34 AGN (see Fig. 6 of Kaspi \et 2000).

\subsection{Optical and X-ray Relation}
\label{discussion.1}

Inspection of Table~\ref{F_var} shows that the X-ray variability
amplitude is about one order of magnitude larger than that of the
optical and is ubiquitous in NLS1s (see e.g. Boller \et 1996, Young \et 1999).
In particular, this result is consistent with the behaviour of NGC 4051
in two previous monitoring campaigns (Done \et 1990, P00). The
striking difference between the variability amplitudes of the X-ray
and the optical bands in NLS1s is not yet understood.

Another interesting result is that each optical emission line varies
about twice as much as its underlying continuum. This trend was also
encountered by Peterson, Crenshaw \& Meyers (1985) and by P00.
One possible reason for the large $F_{var}$ of the lines might be that
the UV and/or the X-ray continua, that are the likely drivers of the
line flux, are varying with much larger amplitudes. However, in 1998 the
H$\beta$ flux remained unchanged when the X-ray source almost completely
turned off and so the highly variable X-ray continuum does not contribute
significantly to the Balmer lines production (P00). Large amplitude UV
variations do, however, remain a possibility and would be consistent with the
very large EUV variations observed by Uttley \et (2000). The other remaining,
and more likely, possibility is that the host-galaxy contribution to the optical
continuum is larger than estimated here ($\sim50\%$; see e.g. the case of
NGC~5548, Gilbert \& Peterson 2003).

\begin{figure}
\centerline{\epsfxsize=2.8in\epsfbox{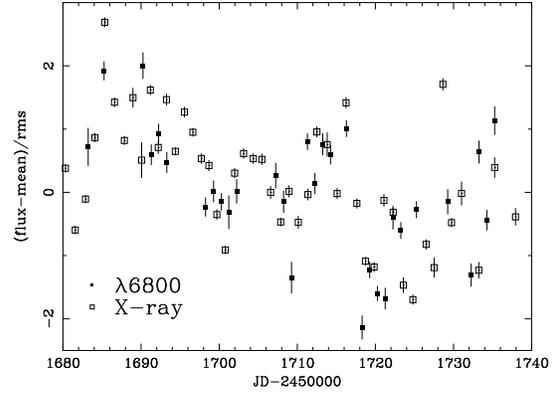}}
\caption{Comparison of X-ray 2--10~keV (open squares) and optical
$\lambda$6800 (filled squares) light curves, renormalised by their
respective rms variability (after mean subtraction).}
\label{rescaledlc}
\end{figure}

The X-ray and optical light curves are apparently correlated at $>95$\%
confidence, although the light curves are not simply correlated which
would lead to a much clearer peak in the CCF (e.g. see the
$\lambda$5100-$\lambda$6800 correlation). The relation between the X-ray and
$\lambda$6800 light curves can be seen by rescaling both light curves (after
subtracting their respective means) by their rms variability (i.e.
$F_{\rm var}$ multiplied by mean flux), as shown in Figure~\ref{rescaledlc}.
No lag is introduced into any light curve. The X-ray and optical light curves
can be seen to be generally correlated, at least on long time-scales, but there
are occasional large discrepancies between the two (which are not attributable
to observational noise) which reduce the strength of the observed correlation.
The differences between the rescaled light curves
in Figure~\ref{rescaledlc} may be attributable to the large
amplitude of short-time-scale X-ray variability relative to the amplitude of
long-time-scale variations,
which implies that the X-ray power spectrum is flatter than the optical power
spectrum. Much better signal-to-noise and sampling in both X-ray and optical
light curves would be required in order to tell if corresponding short-term
variations appear (but at a much weaker level) in the optical light curve.

We find that the peak of the XO cross-correlation is
at zero lag but that the centroid of the CCF lies at an
optical-to-X-ray lag of $2.4\pm1.0$ days.
From simulations we have shown that the probability of exceeding
the observed peak cross-correlation coefficient in random
data from suitably constructed light curves is 4\%.

This result should be compared with two previous attempts to measure
XO lags in NGC~4051. The first attempt (Done \et 1990) found very little
variability ($F_{var}<1$\%) in the optical band in an observational period of
a week (and hence no measurable lag) and the second attempt found a
good correlation on long time-scales with
approximately zero lag (P00). The result of Done et al. (1990) is consistent with our
observations as we find that NGC~4051 shows only small-amplitude optical variability
on time-scales of a week (see Figure~\ref{lc}).  Peterson et al. (2000) were
unable to find any XO correlation on short time-scales, probably due
to relatively poor X-ray sampling compared to
the light curve we present here.  On long time-scales, their
light curves were smoothed by a 30-day boxcar, thus suppressing any rapid X-ray
variations and rendering the detection of a short lag, such as the one
we mention here, impossible.

Also of significant interest is the recent result of Mason \et
(2002). In a 130 ks observation of NGC~4051 with {\sl XMM}-Newton, they
found that the 0.1-12 keV X-ray continuum led the $\lambda2000$ UV
continuum, measured with the {\sl XMM}-Newton optical monitor, by 0.17 days.
The significance of that result is similar to that found here. They
interpret their observation as optical variability arising from
reprocessing of X-ray photons in a region surrounding the central X-ray source.

The fact that the peak of our XO CCF is at zero lag is quite consistent
with the result of Mason \et (2002). Their short observation is not
sensitive to the longer timescales which we sample and we are not able
to resolve the very short timescales which they sample.

Interestingly, we also find that the XO CCF appears to be
asymmetric, in that although it has a zero-lag peak, it has a negative
centroid lag, $\tau_{\rm cent}$ (i.e. X-rays lag optical). A negative
centroid lag and zero peak lag can be reconciled if the variations on
timescales longer than a few hours have a different lag, and hence
probably a different physical origin, than those on shorter timescales.
We can use Monte Carlo simulations to test whether such an observed
centroid lag would be expected by chance from perfectly correlated
light curves (i.e. light curves with true peak and centroid lag of
zero) by counting the number of simulated CCFs with a greater
XO lag than observed\footnote{The light curves are generated
as in \S~\ref{results.2}, but using identical random number
sequences for light curve generation before resampling and applying
Gaussian noise.}.  We find that only 2\% of simulated pairs of
perfectly correlated light curves show centroid lags of greater than
2.4~d, suggesting that the centroid lag and hence observed asymmetry
in the CCF is real (as implied by the lag error estimated using the
FR/RSS method).  However, we caution that the significance of the
centroid lag estimated by Monte Carlo simulations is model-dependent:
we have only tested the significance of the centroid lag assuming
perfectly correlated light curves.

Our putative lag, deduced from the offset position of the centroid of
the CCF might, at first thought, be assumed to result from Compton
upscattering of UV/optical seed photons to produce X-ray photons (e.g.
Haardt \& Maraschi 1991, 1993). A similar argument was used by Uttley
\et (2000) for NGC~4051 to explain the very strong correlation
between the X-ray and EUV emissions, whose variations are simultaneous
to within 1 ks. In that case the size of the X-ray emitting region
was calculated to be $\leq20R_{g}$. Taking account of the increased
spectral difference between the optical and X-ray bands compared to
that between the EUV and X-ray bands, and the greater length of the
putative lag here, the implied size of the X-ray emitting region would
be $\sim1000R_{g}$, much larger than deduced previously. It is hard
to reconcile such a large size with the rapid X-ray variability. It
is also not easy to reconcile an accretion disc of this size (or the
hot corona of such a disc) with the observed properties of this source.

An alternative explanation of an X-ray lag is that the optical
emitting region is further out in the accretion disk than the X-ray
emitting region (which may be a hot corona $<20R_{g}$
in size, Uttley \et 2000).
Variations propagating inwards, perhaps at the
viscous or diffusion timescale (which in NGC~4051 can be quite short, especially
if the disk is thick), would first affect the optical emitting
region and, later, the X-ray emitting region. A similar explanation, based
on the model for flickering in X-ray binaries suggested by
Lyubarskii (1997), is used by Kotov, Churazov and
Gilfanov (2001) to explain the energy dependence of the time lags in the
X-ray variations in Cyg X-1.  Given the many similarities in
variability properties of AGN and X-ray binaries (e.g. Uttley \et 2002, M$^{\rm
c}$Hardy et al., in preparation), such a model might
also be applicable to explain the possible X-ray lag in NGC~4051 .

\section{Conclusions}
\label{conclusions}

Our main conclusions are summarized as follows:

\begin{itemize}
\item
[1.] Variability amplitudes of our X-ray and optical light curves are
consistent with previous observations of NGC~4051. However, despite several
good arguments made to explain the observed emission-line variability
amplitudes, that are larger by a factor of 2 than the optical continuum
variability amplitude, a lack of quantitative evidence remains.

\item
[2.] Our measured $R_{BLR}$ value is $3.0\pm1.5$ light days, which is about
a factor of 2 lower than previous measurements. This implies
M$_{BH}=5^{+6}_{-3}\times10^5 M_{\odot}$, and places NGC~4051 much
closer than before to the best-fit $R_{BLR}$--$L$ slope of Kaspi \et
(2000). The apparent change in BLR distance is not a luminosity effect,
but rather an observational one, since at least the optical flux of the galaxy
remained practically constant during all the monitoring campaigns.

\item
[3.] There is significant evidence for an X-ray/optical
correlation close to zero lag (within one day) in NGC~4051. There is also
evidence that part of the optical flux varies in advance of the X-ray flux by
about 2 days. Although the amplitude of the optical variations is very low,
these observations are consistent with X-ray/optical variations seen
elsewhere and are probably best explained by a combination of effects
including reprocessing of X-ray photons and a physical separation of
the main X-ray and optical producing regions. Although Compton
up-scattering of optical photons to produce X-ray photons cannot be
ruled out, optical photons do not appear to be as important to the
seed photon continuum as UV photons.
\end{itemize}

\section*{Acknowledgments}

We are grateful to WO staff members Ezra Mashal, Friedel Loinger, Sammy Ben-Guigui,
Peter Ibbetson and John Dann for their crucial contribution to this project.
We gratefully acknowledge fruitful discussions with Dan Maoz and Shai Kaspi, and
thank an anonymous referee for useful comments. This work is supported by the Israel
Science Foundation grant 545/00.

\label{lastpage}

\end{document}